\documentclass[a4paper,11pt]{llncs}
\begin{document}
\author{Amit K Awasthi$^1$ \and Sunder Lal$^2$}
\title{A New Proxy Ring Signature Scheme}
\institute {Hindustan College of Science and Technology,\\ Farah, Mathura - 281122 UP  - INDIA\\
\email{awasthi\_hcst@yahoo.com} \and
Institute of Basic Sciences,\\ Khandari, Agra - 282002(UP) - INDIA\\}
\maketitle
\begin{abstract} The concept of ring signature was introduced by Rivest, Shamir and Tauman.
This signature is considered to be a simplified group signature from which identity of signer
is ambiguous although verifier knows the group to which signer belong. In this paper we introduce
a new proxy ring signature scheme.\footnote{This paper appeared at the conference 'RMS 2004', Agra, INDIA}
\end{abstract}

\section{Introduction}
Consider the following situation discussed by Zhang et al. in \cite{Zhangproxyring}-
\begin{quote}An entity delegate his signing capability to many proxies, called proxy
signers set. Any proxy signer can perform the signing operation of the original
entity. These proxy signers want to sign messages on behalf of the original entity
while providing anonymity. Of course, this problem can be solved by group signature
(Take the group manager as the original entity). But in some applications, it is
necessary to protect the privacy of participants (we believe that unconditional
anonymity is necessary in many occasions). If the proxies don't hope that some one
(include the original signer) can open their identities, the group signature is not
suitable in here (Because a group manager can open the signature to reveal the
identity of the signer). \end{quote} To solve above problem proxy ring signature
scheme was introduced by Zhang et al \cite{Zhangproxyring}. In this paper, we propose
a new proxy ring signature scheme based on Lin-Wu's \cite{newringsigscheme} ID-based
ring signature scheme from bilinear pairing. As compared to previous scheme
\cite{Zhangproxyring}, our scheme is computationally more efficient, especially for
the pairing operations required  during signature verification.

\section{Preliminaries}
\begin{itemize}
\item \textbf{Bilinear Pairings}\\
    Let ${\textbf{G}}_{1}$ cyclic additive group generated by $P$, whose order is a prime q,
and G2 bea cyclic multiplicative group of the same order q: A bilinear pairing is a
map e : ${\textbf{G}}_{1} \times {\textbf{G}}_{1}\longrightarrow {\textbf{G}}_{2}$
with the following properties:

P1: \textit{Bilinear}: $e(aP, bQ) = {e(P,Q)}^{ab}$;

P2: \textit{Non-degenerate}: There exists $P,Q \in {\textbf{G}}_{1}$ such that $e(P,Q)
\neq 1$;

P3: \textit{Computable}: There is an efficient algorithm to compute $e(P,Q)$ for all
$P, Q \in {\textbf{G}}_{1}$.

    When the DDHP (Decision Diffie-Hellman Problem) is easy but the CDHP (Computational
Diffie-Hellman Problem) is hard on the group $G$; we call $G$ a Gap Diffie-Hellman
(GDH) group. Such groups can be found on supersingular elliptic curves or
hyperelliptic curves over finite field, and the bilinear parings can be derived from
the Weil or Tate pairing. We can refer to \cite{{Boneh},{Zhangproxyring}} for more
details.

Through this paper, we define the system parameters in all schemes are as follows: Let
$P$ be a generator of ${\textbf{G}}_{1}$; the bilinear pairing is given by $ e :
{\textbf{G}}_{1} \times {\textbf{G}}_{1}\longrightarrow {\textbf{G}}_{2}$. Define two
cryptographic hash functions ${H}_{1} : {\{0, 1\}}^{*} \longrightarrow {z}_{q}$ and
${H}_{2} : {\{0, 1\}}^{*} \longrightarrow {\textbf{G}}_{2} $.

\item {\textbf{Paring based short signature scheme (PBSSS)}} Boneh et al.'s pairing
based short
signature scheme is as follows-\\
    \textbf{Key Generation} Choose a random number $s \in {Z}_{q}^{*}$ and compute
    ${P}_{Pub} = sP$ where $({G}_{1},{G}_{2},q,P,{P}_{Pub},{H}_{2})$ are public
    parameter and $s$ is the secret key.\\\textbf{Signing} For a message $m \in {{0,1}}^{*}$, compute ${P}_{m} = {H}_{2}(m) \in
    {G}_{1},$. Then ${S}_{m} = s{P}_{m}$ is the signature on message $m$.\\\textbf{Verification} Check whether the following equation holds
    \begin{center}
$ e({S}_{m}, P) = e({H}_{2}(m), {P}_{Pub})$
    \end{center}
This scheme is proven to be secure against existential forgery on adaptive chosen
message attack (in random oracle model) assuming Computationally Diffie Hellman
Problem is hard \cite{boneh}.
\end{itemize}
\section{Proposed Scheme}

\textbf{[Setup]} The system parameters \verb"params" = \{${\textbf{G}}_{1},
{\textbf{G}}_{2},e ,q,P,{H}_{1}, {H}_{2}\}$ Let Alice be the original signer with
public key ${PK}_{o} = {s}_{o} P$ and private key ${s}_{o}$, and $L = \{{PS}_{i}\}$ be
the set of proxy signers with public key ${PK}_{{p}_{i}} = {s}_{{p}_{i}} P$ and
private key ${s}_{{p}_{i}}$.\\\textbf{[Proxy Key Generation]} The original signer
prepares a warrant $w$, which is explicit description of the delegation relation. Then
he sends $(w, {s}_{o} {H}_{2}(w))$ to the proxy group L. Each proxy signer uses his
secret key ${S}_{{p}_{i}}$ to sign the warrant $w$ and gets his proxy key ${S}_{i} =
{s}_{o} {H}_{2}(w) + {s}_{{p}_{i}} {H}_{2}(w)$.\\\textbf{[Proxy Ring Signing]}For
signing any message $m$, the proxy signer ${PS}_{i}$ chooses a subset $L\prime
\subseteq L$. Proxy signers's public key is listed in ${L}_{\prime}$. Now to sign he/
she perform following operations:

\begin{itemize}
\item{Initialization}: Choose randomly an element $A \in {\textbf{G}}_{1}$, compute
\begin{equation}
{c}_{k+1} = e(A, P)
\end{equation}
%
\item{Generate forward ring sequence} For $i = k + 1,k + 2,....k+(n -1)$ choose
randomly ${T}_{i} \in {\textbf{G}}_{1}$ and compute
\begin{equation}
{c}_{i+1} = e{({PK}_{o}+{PK}_{{p}_{i}}, {c}_{i}{H}_{2}(w))}^{{H}_{2}(m \parallel L)}.
e(Ti, P)
\end{equation}
\item{Forming the ring}: Let ${R}_{n}={R}_{o}$. Then, ${PS}_{i}$ computes
\begin{equation}
{T}_{i} = A - {h}_{2}(m \parallel L) {c}_{i}{S}_{i},
\end{equation}
\begin{equation}
T = {\Sigma}_{i=1}^{n} {T}_{i}
\end{equation}
\item{Output}: Finally, Let ${c}_{n} = {c}_{0}$. The resulting ring signature for a
message $m$ and with ring member specified by $L \prime$ is the
$(n+1)$-tuple:$({c}_{1}, {c}_{2}, ..., {c}_{n}, T)$
\end{itemize}
\paragraph{}
\textbf{[Verification]} Given message $m$, its ring signature $({c}_{1}, {c}_{2}, ...,
{c}_{n}, T)$, and the set $L \prime$ of the identities of all ring members, the
verifier can check the validity of the signature by the testing if:

\begin{equation}
{\Pi}_{i=1}^{n} {c}_{i} = e{({PK}_{o}+{PK}_{{p}_{i}}, {\Sigma}_{i=1}^{n} {c}_{i}
{H}_{1}(w))}^{{H}_{2}(m \parallel L)}.e(T,P)
\end{equation}

\section{Performance and Security Discussion}

\textbf{Key Secrecy} In computing user ${P}_{i}$'s private key ${S}_{i}$ from the
corresponding public key ${PK}_{o} +{PK}_{{P}_{i}}$ requires the knowledge of original
signer's private key ${s}_{o}$ and proxy signer's private key ${s}_{{p}_{i}}$.
According to definition these keys are protected under the intractability of DLP in
${G}_{1}$ as ${PK}_{o}= {s}_{o}P$ and ${PK}_{{P}_{i}} = {s}_{{p}_{i}}P$.\\
\textbf{Signer ambiguity} In a valid proxy ring signature $({c}_{1}, {c}_{2}, ...,
{c}_{n}, T)$ with proxy group ${L}_{\prime}$ generated by ${PS}_{i}$ all ${c}_{i}$'s
are computed by eq 2. Since ${T}_{i} \in {G}_{1}$ is chosen uniformly at random, each
${c}_{i}$ is uniformly distributed over ${G}_{2}$. Thus, regardless who the actual
signer is and how many ring members involved $({c}_{1}, {c}_{2}, ..., {c}_{n})$ biases
to no specific ring member.

\section{Conclusion}In this paper we proposed a new proxy ring signature which becomes
needful whenever proxy signer want to sign message on behalf of the original signer
providing anonymity. The proposed scheme is more efficient as compared with the Zhang
et al.'s, especially for the pairing operation required in the signature verification.
This proxy ring signature scheme is more efficient for those verifiers who have
limited computing power.


\begin{thebibliography}{4}

\bibitem{Mam1}
Mambo, M., Usuda, K., Okamoto, E.: Proxy Signature: Delegation of the power to sign
messages, IEICE Trans. Fundamentals, E79--A:9, (1996) 1338 - 1353 5.
%
\bibitem{Mam2}
Mambo, M., Usuda, K., Okamoto, E.: Proxy signatures for delegating signing operation,
Proc. 3rd ACM Conference on Computer and Communication Security, (1996) 48--57
%
\bibitem{Kim1}
Kim, S., Park, S., Won, D.: Proxy signatures: Revisited, ICICS'97, LNCS Vol.~1334,
Springer-Verlag, (1997) 223--232
%
\bibitem{Lee1}
Lee, W. B., Chang, C. Y.: Efficient Proxy-protected proxy signature scheme based on
discrete logarithm, Proceedings of 10th Conference on Information Security, Hualien,
Taiwan, ROC, (2000) 4-7
%
\bibitem{Abe}
M. Abe, M. Ohkubo, and K. Suzuki, 1-out-of-n signatures from a variety of keys, To
appear in Advances in Cryptology-Asiacrypt 2002, 2002.
\bibitem{Barr}
P.S.L.M. Barreto, H.Y. Kim, B.Lynn, and M.Scott, Effient algorithms for pairing- based
cryptosystems, Advances in Cryptology-Crypto 2002, LNCS 2442, pp.354- 368,
Springer-Verlag, 2002.
\bibitem{Boneh}
D. Boneh and M. Franklin, Identity-based encryption from the Weil pairing, Advances in
Cryptology-Crypto 2001, LNCS 2139, pp.213-229, Springer-Verlag, 2001.
\bibitem{boneh}
D. Boneh, B. Lynn, and H. Shacham, Short signatures from the Weil pairing, Advances in
Cryptology-Asiacrypt 2001, LNCS 2248, pp.514-532, Springer-Verlag, 2001.
\bibitem{newringsigscheme}
C. Y. Lin and T. C Wu, An Identity-based Ring Signature Scheme form Bilinear
Pairings,Cryptology ePrint Archive, Report 2003/117 available at
http://eprint.iacr.org/2003/117/.
\bibitem{Zhangproxyring}
F. Zhang, R. S. Naini and C Y Lin, New Proxy Signature, Proxy Blind Signature and
Proxy Ring Signature Schemes from Bilinear Pairings,Cryptology ePrint Archive, Report
2003/117 available at http://eprint.iacr.org/2003/104/.
\bibitem{ZhangRingSS}
F. Zhang and K. Kim, ID-based Blind Signature and Ring Signature from Pairings,
Advances in Cryptology, Asiacrypt 2002, LNCS 2501, Springer-Verlag, pp 533-547.
\end{thebibliography}
\end{document}